\documentstyle[aps]{revtex}

\begin{document}
\title{On entanglement - assisted classical capacity}
\author{A.~S. Holevo}
\address{Steklov Mathematical Institute, Gubkina 8, 117966 Moscow, Russia}

\maketitle
\begin{abstract}
We give a modified proof of the recent result of C. H. Bennett, P. W. Shor, J. A. Smolin and A. V.
Thapliyal concerning entanglement-assisted classical capacity of
quantum channel and discuss relation between  entanglement-assisted and
unassisted classical capacities. \end{abstract}
\newpage
\section{INTRODUCTION}
The classical capacity of a quantum channel is the capacity of
transmission of classical information through the channel. It is well
known that the classical capacity can be increased if there is an
additional resource in the form of entangled state shared between the
input $A$ and the output $B$ of the channel. While entanglement itself cannot
serve for transmission of information from $A$ to $B$, it may enhance the
transmission provided there is a quantum channel connecting the systems.
If the channel is ideal (i. e. the identity map Id from $A$ to $B$) then the
entanglement-assisted capacity is twice as great as the unassisted
classical capacity, the enhancement being realized by the dense coding
protocol \cite{wie}.

Recently C. H.  Bennett, P.  W. Shor, J. A.  Smolin and A.  V.
Thapliyal (BSST) studied the effect of the shared entanglement on the
classical capacity of quantum non-ideal (noisy) channel
\cite{eac,eac2} and  obtained a remarkably simple formula for the
entanglement-assisted capacity in terms of the maximal mutual quantum
information between $A$ and $B$. The proof, which is by no means trivial,
involves in particular a rather tricky derivation of an important continuity
property of quantum entropy. In this paper we give a modified proof of the
BSST theorem, including a more transparent proof of this property; moreover,
we make further simplifications by using heavily properties of conditional
quantum entropy rather than the underlying strong subadditivity.

In Ref. \cite{eac} it was shown that the enhancement in the classical
 capacity can achieve arbitrarily large values. To this end the case of
$d$-dimensional depolarizing channel in the limit of strong noise
($p\rightarrow 1$) was considered; we remark that the enhancement is even
greater for the extreme case $p=\frac{d^2}{d^2-1}$. Moreover, we derive a
general inequality between entanglement-assisted and unassisted capacities
which may be relevant to the additivity problem in quantum information
theory.

\section{THE BSST THEOREM}
We refer the reader to Refs. \cite{nc,hw} for some basic definitions and
 results of quantum information theory used in this paper.

 Consider the following protocol for
the classical information transmission through a quantum channel $\Phi $.
Systems $A$ and $B$ of the same dimension share an entangled (pure) state
$S_{AB}$.  $A$ does some encoding $ i\rightarrow {\mathcal E}_{A}^{i}$
depending on a classical signal $i$ with probabilities $\pi _{i}$ and sends
its part of this shared state through the channel $\Phi $ to $B$. Thus $B$
gets the states $(\Phi \otimes {\mathrm Id} _{B})\left[ S_{AB}^{i}\right] ,$
where $S_{AB}^{i}=({\mathcal E} _{A}^{i}\otimes {\mathrm Id}_{B})\left[
S_{AB}\right] ,$ with probabilities $ \pi _{i},$ and $B$ is trying to extract
the maximal classical information by doing measurements on these states. This
is similar to the dense coding, but instead of the ideal channel, $A$ uses a
noisy channel $\Phi $. We now look for the classical capacity of this
protocol, which is called \textit{ entanglement-assisted classical capacity}
of the channel $\Phi $.

The maximum over measurements of $B$ can be evaluated using the coding
theorem for the classical capacity \cite{hol}. First we have the \textit{\
one-shot entanglement-assisted classical capacity}
\begin{equation}
C_{ea}^{(1)}(\Phi )=\max_{\pi _{i},{}{\mathcal E}_{A}^{i},S_{AB}}\left[
H\left( \sum_{i}\pi _{i}(\Phi \otimes {\mathrm Id}_{B})\left[ S_{AB}^{i}
\right] \right) -\sum_{i}\pi _{i}H\left( (\Phi \otimes {\mathrm Id} _{B})
\left[ S_{AB}^{i}\right] \right) \right] ,
\end{equation}
where $H(S)$ denotes the von Neumann entropy of the density operator $S$.
Using the channel $n$ times and allowing entangled measurements on $B$'s
side, one gets
\begin{equation}
C_{ea}^{(n)}(\Phi )=C_{ea}^{(1)}(\Phi ^{\otimes n}).
\end{equation}
The full entanglement-assisted classical capacity is then
\begin{equation}
C_{ea}(\Phi )=\lim_{n\rightarrow \infty }\frac{1}{n}C_{ea}^{(1)}(\Phi
^{\otimes n}).
\end{equation}

The following result was announced in Ref. \cite{eac}, and a proof was given
in Ref. \cite{eac2}:  \begin{equation} C_{ea}(\Phi )=\max_{S_{A}}I(S_{A};
\Phi ), \label{thm} \end{equation} where \begin{equation} I(S_{A}; \Phi
)=H(S_{A})+H(\Phi (S_{A}))-H(S_{A};\Phi )  \label{qmi} \end{equation} is the
quantum mutual information, with $H(S_{A};\Phi )$ denoting the entropy
exchange (see Refs. \cite{nc,hw}). Below we give a simplified proof of this
remarkable formula.

\textit{Proof} of the inequality
\begin{equation}
C_{ea}(\Phi )\geq \max_{S_{A}}I(S_{A}; \Phi ).  \label{geq}
\end{equation}
It is shown in \cite{eac2} by generalizing the dense coding protocol that
\begin{equation}
C_{ea}^{(1)}(\Phi ^{\otimes n})\geq I\left( \frac{P}{\dim P},\Phi ^{\otimes
n}\right)  \label{proj}
\end{equation}
for arbitrary projection $P$ in ${\mathcal H}_{A}^{\otimes n}.$
We give this proof for completeness here.
Indeed, let $
P=\sum_{k=1}^{m}|e_{k}\rangle \langle e_{k}|,$ where $\left\{
e_{k};k=1,\dots ,m=\dim P\right\} $ is an orthonormal system. Define
unitary operators in ${\mathcal H}_{A}$ acting as
\begin{eqnarray*}
V|e_{k}\rangle &=&\exp \left( \frac{2\pi ik}{m}\right) |e_{k}\rangle ;\quad
U|e_{k}\rangle =|e_{k+1({\mathrm mod}m)}\rangle ;\quad k=1,\dots ,m, \\
W_{\alpha \beta } &=&U^{\alpha }V^{\beta };\quad \alpha ,\beta =1,\dots ,m
\end{eqnarray*}
on the subspace generated by $\left\{ e_{k}\right\} ,$ and as the identity
onto its orthogonal complement. The operators $W_{\alpha \beta }$ are a
finite-dimensional version of the Weyl-Segal operators for Boson systems
(see e. g. \cite{hw}).
Let
\[
|\psi _{AB}\rangle =\frac{1}{\sqrt{m}}\sum_{k=1}^{m}|e_{k}\rangle \otimes
|e_{k}\rangle .
\]
Then it is  easy  to show that

1) $\left( W_{\alpha \beta }\otimes I_{B}\right) |\psi _{AB}\rangle ;\quad
\alpha ,\beta =1,\dots ,m,$ is an orthonormal system in ${\mathcal H}
_{A}\otimes {\mathcal H}_{B}$ ; in particular, if $m=\dim {\mathcal H}_{A}$,
it is a basis;

2) $\sum_{\alpha ,\beta =1}^{m}\left( W_{\alpha \beta }\otimes I_{B}\right)
|\psi _{AB}\rangle \langle \psi _{AB}|\left( W_{\alpha \beta }\otimes
I_{B}\right) ^{\ast }=P\otimes P.$

Thus operators $\left\{ W_{\alpha \beta };\quad \alpha ,\beta =1,\dots
,m\right\} $ play a role similar to Pauli matrices in the dense coding
protocol for qubits.

Take the classical signal to be transmitted as $i=(\alpha ,\beta )$ with
equal probabilities $1/m^{2},$ the entangled state $|\psi _{AB}\rangle
\langle \psi _{AB}|$ and the unitary encodings ${\mathcal E}_{A}^{i}\left[ S
\right] =W_{\alpha \beta }SW_{\alpha \beta }^{\ast }.$ Then we have
\[
C_{ea}^{(1)}(\Phi ^{\otimes n})\geq H\left( \frac{1}{m^{2}}\sum_{\alpha
\beta }(\Phi \otimes {\mathrm Id}_{B})[S_{AB}^{\alpha \beta }]\right) -\frac{1
}{m^{2}}\sum_{\alpha \beta }H\left( (\Phi \otimes {\mathrm Id}
_{B})[S_{AB}^{\alpha \beta }]\right) ,
\]
where $S_{AB}^{\alpha \beta }=\left( W_{\alpha \beta }\otimes I_{B}\right)
|\psi _{AB}\rangle \langle \psi _{AB}|\left( W_{\alpha \beta }\otimes
I_{B}\right) ^{\ast }.$ Then by the property 2) the first term in the right
hand side is equal to $H\left( (\Phi \otimes {\mathrm Id}_{B})\left[ \frac{P}{
m}\otimes \frac{P}{m}\right] \right) =H\left( \frac{P}{m}\right) +H\left(
\Phi \left[ \frac{P}{m}\right] \right) .$ Since $S_{AB}^{\alpha \beta }$ is
a purification of $\frac{P}{m}$ in ${\mathcal H}_{B},$ the entropies in the
second term are all equal to $H\left( \frac{P}{m},\Phi \right) .$ By the
expression for quantum mutual information (\ref{qmi}) this proves (\ref{proj}).
For future use, note that the last term in the quantum mutual information
-- the entropy exchange $H(S_{A}; \Phi )$-- is equal to the final environment
entropy $H(\Phi _{E}\left[ S_{A}\right] ),$ where $\Phi _{E}$ is a channel
from the system space ${\mathcal H}_{A}$ to the environment space ${\mathcal H}
_{E}$ the actual form of which we need not to know (see \cite{hw}).

Now let $S_{A}=S$ be an arbitrary state in ${\mathcal H}_{A},$ and let $
P^{n,\delta }$ be the typical projection of the state $S^{\otimes n}$ in $
{\mathcal H}_{A}^{\otimes n}.$ It was suggested in \cite{eac} that for
\textit{arbitrary} channel $\Psi $ from ${\mathcal H}_{A}$ to possibly other
Hilbert space $\tilde{{\mathcal H}}$
\[
\lim_{\delta \rightarrow 0}\lim_{n\rightarrow \infty }\frac{1}{n}H\left(
\Psi ^{\otimes n}\left( \frac{P^{n,\delta }}{\dim P^{n,\delta }}\right)
\right) =H(\Psi (S)),
\]
which would imply, by the expressions for the mutual information and the
entropy exchange, that
\begin{equation}
\lim_{\delta \rightarrow 0}\lim_{n\rightarrow \infty }\frac{1}{n}I\left(
\frac{P^{n,\delta }}{\dim P^{n,\delta }}; \Phi ^{\otimes n}\right) =I(S; \Phi
),  \label{lim}
\end{equation}
and hence, by (\ref{proj}), the required inequality (\ref{geq}). We shall
prove (\ref{lim}) with $P^{n,\delta }$ being the \textit{strongly typical
projection} of the state $S^{\otimes n}.$

Let us fix small positive $\delta $, and let $\lambda _{j}$ be the
eigenvalues, $|e_{j}\rangle $ the eigenvectors of the density operator $S$.
Then the eigenvalues and eigenvectors of ${S}^{\otimes n}$ are $\lambda
_{J}=\lambda _{j_{1}}\cdot ...\cdot \lambda _{j_{n}},\quad |e_{J}\rangle
=|e_{j_{1}}\rangle \otimes ...\otimes |e_{j_{n}}\rangle $ where $
J=(j_{1},...,j_{n})$. The sequence $J$ is called \textit{strongly typical }
\cite{czi} if the numbers $N(j|J)$ of appearance of the symbol $j$ in $J$
satisfy the condition
\[
\left| \frac{N(j|J)}{n}-\lambda _{j}\right| <\delta ,\quad j=1,\dots ,d,
\]
and $N(j|J)=0$ if $\lambda _{j}=0.$ Let us denote the collection of all
strongly typical sequences as $B^{n,\delta },$ and let $\mathsf{P}^{n}$ be
the probability distribution given by the eigenvalues $\lambda _{J}.$ Then
by the Law of Large Numbers, $\mathsf{P}^{n}\left( B^{n,\delta }\right)
\rightarrow 1$ as $n\rightarrow \infty .$ It is shown in \cite{czi} that the
size of $B^{n,\delta }$ satisfies
\begin{equation}
2^{n[H(S)-\Delta _{n}(\delta )]}<\left| B^{n,\delta }\right|
<2^{n[H(S)+\Delta _{n}(\delta )]},  \label{size}
\end{equation}
where $H({S})=-\sum_{j=1}^{d}\lambda _{j}\log \lambda _{j},$ and $
\lim_{\delta \rightarrow 0}\lim_{n\rightarrow \infty }\Delta _{n}(\delta
)=0. $

For arbitrary function $f(j),j=1,\dots ,d,$ and $J=(j_{1},...,j_{n})\in
B^{n,\delta }$ we have
\begin{equation}
\left| \frac{f(j_{1})+\dots +f(j_{n})}{n}-\sum_{j=1}^{d}\lambda
_{j}f(j)\right| <\delta \max f.  \label{lln}
\end{equation}
In particular, any strongly typical sequence is (entropy) typical: taking $
f(j)=-\log \lambda _{j}$ gives
\begin{equation}
n[H({S})-\delta _{1}]<-\log \lambda _{J}<n[H(S)+\delta _{1}],  \label{ent}
\end{equation}
where $\delta _{1}=\delta \max_{\lambda _{j}>0}(-\log \lambda _{j}).$ The
converse is not true -- not every typical sequence is strongly typical.

The strongly typical projector is defined as the following spectral
projector of $S^{\otimes n}$:
\[
P^{n,\delta }=\sum_{J\in B^{n,\delta }}|e_{J}\rangle \langle e_{J}|.
\]
We denote $d_{n,\delta }=\dim P^{n,\delta }=\left| B^{n,\delta }\right| $
and $\bar{S}^{n,\delta }=\frac{P^{n,\delta }}{d_{n,\delta }}$ and we are
going to prove that
\begin{equation}
\lim_{\delta \rightarrow 0}\lim_{n\rightarrow \infty }\frac{1}{n}H(\Psi
^{\otimes n}\left( \bar{S}^{n,\delta }\right) )=H(\Psi (S))  \label{llim}
\end{equation}
for arbitrary channel $\Psi $.

We have
\[
nH(\Psi (S))-H(\Psi ^{\otimes n}\left( \bar{S}^{n,\delta }\right) ) =H(\Psi
(S)^{\otimes n})-H(\Psi ^{\otimes n}\left( \bar{S}^{n,\delta }\right) )
\]
\begin{equation}
=H\left( \Psi ^{\otimes n}\left( \bar{S}^{n,\delta }\right) |\Psi ^{\otimes
n}(S^{\otimes n})\right) +{\mathrm Tr\log }\Psi (S)^{\otimes n}\left( \Psi
^{\otimes n}\left( \bar{S}^{n,\delta }\right) -\Psi (S)^{\otimes n}\right) ,
\label{main}
\end{equation}
where $H\left( \cdot |\cdot \right) $ is relative entropy. Strictly
speaking, this formula is correct if the density operator $\Psi (S)^{\otimes
n}$ is nondegenerate, which we assume for a moment. Later we shall show how
the argument can be modified to the general case.

For the first term we have the estimate by the fundamental
property of monotonicity of the relative entropy \[ H\left( \Psi ^{\otimes
n}\left( \bar{S}^{n,\delta }\right) |\Psi ^{\otimes n}(S^{\otimes n})\right)
\le H\left( \bar{S}^{n,\delta }|S^{\otimes n}\right) , \] with the right hand
side computed explicitly as \[ H\left( \bar{S}^{n,\delta }|S^{\otimes
n}\right) =\sum_{J\in B^{n,\delta }} \frac{1}{d_{n,\delta }}\log
\frac{1}{d_{n,\delta }\lambda _{J}}=-\log d_{n,\delta }-\sum_{J\in
B^{n,\delta }}\frac{1}{d_{n,\delta }}\log \lambda _{J}, \] which is less than
or equal to $n\left( \delta _{1}+\Delta _{n}(\delta )\right) $ by
(\ref{ent}), (\ref{size}), giving sufficient estimate.

By using the identity
\[
{\mathrm \log }\Psi (S)^{\otimes n}={\mathrm \log }\Psi (S)\otimes I\otimes
\dots \otimes I+\dots +I\otimes \dots \otimes I\otimes {\mathrm \log }\Psi
(S),
\]
and introducing the operator $F=\Psi ^{*}({\mathrm \log }\Psi (S))$ where $
\Psi ^{*}$ is the dual channel, we can rewrite the second term as
\begin{eqnarray*}
&&n{\mathrm Tr}\frac{\left( F\otimes I\otimes \dots \otimes I+\dots +I\otimes
\dots \otimes I\otimes F\right) }{n}\left( \bar{S}^{n,\delta }-S^{\otimes
n}\right) \\
&=&\frac{n}{d_{n,\delta }}\sum_{J\in B^{n,\delta }}\left[ \frac{
f(j_{1})+\dots +f(j_{n})}{n}-\sum_{j=1}^{d}\lambda _{j}f(j)\right] ,
\end{eqnarray*}
where $f(j)=\langle e_{j}|F|e_{j}\rangle ,$ which is evaluated by $n\delta
\max f$ via (\ref{lln}). This establishes (\ref{llim}) in the case of a
nondegenerate $\Psi (S).$

Coming back to the general case, let us denote $P_{\Psi }$ the supporting
projector of $\Psi (S)$. Then the supporting projector of $\Psi (S)^{\otimes
n}$ is $P_{\Psi }^{\otimes n},$ and the support of $\Psi ^{\otimes n}\left(
\bar{ S}^{n,\delta }\right) $ is contained in the support of $\Psi
(S)^{\otimes n}=\Psi ^{\otimes n}(S^{\otimes n}),$ because the support of
$\bar{S} ^{n,\delta }$ is contained in the support of $S^{\otimes n}.$ Thus
the second term in (\ref{main}) should be understood as \[ {\mathrm
Tr}P_{\Psi }^{\otimes n}{\mathrm \log }\left[ P_{\Psi }^{\otimes n}\Psi
(S)^{\otimes n}P_{\Psi }^{\otimes n}\right] P_{\Psi }^{\otimes n}\left( \Psi
^{\otimes n}\left( \bar{S}^{n,\delta }\right) -\Psi (S)^{\otimes n}\right) ,
\] where now we have $\log $ of a nondegenerate operator in $P_{\Psi
}^{\otimes n}{\mathcal H}_{A}^{\otimes n}.$ We can then repeat the argument
with $F$ defined as $\Psi ^{*}(P_{\Psi }\left[ {\mathrm \log }P_{\Psi }\Psi
(S)P_{\Psi }\right] P_{\Psi }).$ This fulfills the proof of (\ref{lim}), from
which (\ref{geq}) follows.

\textit{Proof} of the inequality
\begin{equation}
C_{ea}(\Phi )\leq \max_{S_{A}}I(S_{A},\Phi ).  \label{ine}
\end{equation}
We first prove that
\begin{equation}
C_{ea}^{(1)}(\Phi )\leq \max_{S_{A}}I(S_{A},\Phi ).  \label{eac}
\end{equation}
The proof is a modification of that from \cite{eac2}, using properties of
conditional quantum entropy which are known to follow from the strong
subadditivity of the entropy (see e. g. \cite{wehrl}, \cite{nc}), rather
than the strong subadditivity itself.

Let us denote ${\mathcal E}_{A}^{i}$ the encodings used by $A$. Let $S_{AB}$
be the pure state initially shared by $A$ and $B$, then the state of the
system $AB$ (resp. $A$) after the encoding is
\begin{equation}
S_{AB}^{i}=({\mathcal E}_{A}^{i}\otimes {\mathrm Id}_{B})[S_{AB}],\quad
{\mathrm resp}.{\mathrm \quad }S_{A}^{i}={\mathcal E}_{A}^{i}[S_{A}].
\label{enc}
\end{equation}
Note that the partial state of $B$ does not change after the encoding, $
S_{B}^{i}=S_{B}.$ We are going to prove that
\[
H\left( \sum_{i}\pi _{i}(\Phi \otimes {\mathrm Id}_{B})[S_{AB}^{i}]\right)
-\sum_{i}\pi _{i}H\left( (\Phi \otimes {\mathrm Id}_{B})[S_{AB}^{i}]\right)
\]
\begin{equation}
\le I\left( \sum_{i}\pi _{i}S_{A}^{i};\Phi \right) .  \label{ineq}
\end{equation}
By the quantum coding theorem, the maximum of the left hand side with
respect to all possible $\pi _{i},{\mathcal E}_{A}^{i}$ is just $
C_{ea}^{(1)}(\Phi ),$ whence (\ref{eac}) will follow.

By using subadditivity of quantum entropy, we can evaluate the first term in
the left hand side of (\ref{ineq}) as
\[
H\left( \sum_{i}\pi _{i}\Phi [S_{A}^{i}]\right) +H(S_{B})=H\left( \Phi \left[
\sum_{i}\pi _{i}S_{A}^{i}\right] \right) +\sum_{i}\pi _{i}H(S_{B}).
\]
Here the first term already gives the output entropy from $I\left(
\sum_{i}\pi _{i}S_{A}^{i};\Phi \right) .$ Let us proceed with evaluation of
the remainder
\[
\sum_{i}\pi _{i}\left[ H(S_{B})-H\left( (\Phi \otimes {\mathrm Id}
_{B})[S_{AB}^{i}]\right) \right] .
\]
We first show that the term in squared brackets does not exceed $
H(S_{A}^{i})-H\left( (\Phi \otimes {\mathrm Id}_{R^{i}})[S_{AR^{i}}^{i}]
\right) ,$ where $R^{i}$ is the purifying (reference) system for $S_{A}^{i},$
and $S_{AR^{i}}^{i}$ is the purified state. To this end consider the unitary
extension of the encoding ${\mathcal E}_{A}^{i}$ with the environment $E_{i},$
which is initially in a pure state. From (\ref{enc}) we see that we can take
$R^{i}=BE_{i}$ (after the unitary interaction which involves only $AE_{i}).$
Then, again denoting with primes the states after the application of the
channel $\Phi ,$ we have
\begin{equation}
H(S_{B})-H\left( (\Phi \otimes {\mathrm Id}_{B})[S_{AB}^{i}]\right)
=H(S_{B})-H(S_{A^{\prime }B}^{i})=-H_{i}(A^{\prime }|B),  \label{hren}
\end{equation}
where the lower index $i$ of the conditional entropy points out to the joint
state $S_{A^{\prime }B}^{i}.$ Similarly
\[
H(S_{A}^{i})-H\left( (\Phi \otimes {\mathrm Id}_{R^{i}})[S_{AR^{i}}^{i}]
\right) =H(S_{R^{i}}^{i})-H\left( S_{A^{\prime }R^{i}}^{i}\right)
\]
\[
=-H_{i}(A^{\prime }|R^{i})=-H_{i}(A^{\prime }|BE_{i}),
\]
which is greater or equal than (\ref{hren}) by monotonicity of the
conditional entropy.

Using the concavity of the function $S_{A}\rightarrow H(S_{A})-H\left( (\Phi
\otimes {\mathrm Id}_{R})[S_{AR}]\right) $ to be shown below, we get
\[
\sum_{i}\pi _{i}\left[ H(S_{A}^{i})-H\left( (\Phi \otimes {\mathrm Id}
_{R^{i}})[S_{AR^{i}}^{i}]\right) \right]
\]
\[
\leq H\left( \sum_{i}\pi _{i}S_{A}^{i}\right) -H\left( (\Phi \otimes {\mathrm
\ Id}_{R})[\hat{S}_{AR}]\right) ,
\]
where $\hat{S}_{AR}$ is the state purifying $\sum_{i}\pi _{i}S_{A}^{i}$ with
a reference system $R.$

To complete this proof it remains to show the above concavity. By
introducing the environment $E$ for the channel $\Phi $, we have
\[
H(S_{A})-H\left( (\Phi \otimes {\mathrm Id}_{R})[S_{AR}]\right)
=H(S_{R})-H\left( S_{A^{\prime }R}\right)
\]
\[
=H(S_{A^{\prime }E^{\prime }})-H(S_{E^{\prime }})=H(A^{\prime }|E^{\prime })
\]
The conditional entropy $H(A^{\prime }|E^{\prime })$ is a concave function
of $S_{A^{\prime }E^{\prime }}$. The map $S_{A}\rightarrow S_{A^{\prime
}E^{\prime }}$ is affine and therefore $H(A^{\prime }|E^{\prime })$ is a
concave function of $S_{A}$ .

Applying the same argument to the channel $\Phi ^{\otimes n}$ gives
\begin{equation}
C_{ea}^{(n)}(\Phi )\leq \max_{S_{A}^{n}}I(S_{A}^{n};\Phi ^{\otimes n}).
\end{equation}
Then from subadditivity of quantum mutual information \cite{adami}, we have
\[
\max_{S_{12}}I(S_{12};\Phi _{1}\otimes \Phi _{2})=\max_{S_{1}}I(S_{1};\Phi
_{1})+\max_{S_{2}}I(S_{2};\Phi _{2}),
\]
implying the remarkable additivity property
\[
\max_{S_{A}^{n}}I(S_{A}^{n};\Phi ^{\otimes n})=n\max_{S_{A}}I(S_{A};\Phi ).
\]
Therefore, finally we obtain (\ref{ine}).

\section{RELATION BETWEEN ENTANGLEMENT-ASSISTED AND UNASSISTED CAPACITIES}

The definition of $C_{ea}^{(1)}(\Phi )$ and hence of $C_{ea}(\Phi )$ can be
formulated without explicit introduction of the encoding operations
${\mathcal E}_{A}^{i},$ namely \begin{equation} C_{ea}^{(1)}(\Phi )=\max_{\pi
_{i},{}\left\{ S_{AB}^{i}\right\} \in \Sigma _{B}}\left[ H\left( \sum_{i}\pi
_{i}(\Phi \otimes {\mathrm Id}_{B})\left[ S_{AB}^{i}\right] \right)
-\sum_{i}\pi _{i}H\left( (\Phi \otimes {\mathrm Id} _{B})\left[
S_{AB}^{i}\right] \right) \right] , \end{equation} where $\Sigma _{B}$ is the
collection of families of the states $\left\{ S_{AB}^{i}\right\} $ satisfying
the condition that their partial states $ S_{B}^{i}$ do not depend on $i,$
$S_{B}^{i}=S_{B}.$ This follows from

\textbf{Lemma}. Let $\left\{ S_{AB}^{i}\right\} $ be a family of the states
satisfying the condition $S_{B}^{i}=S_{B}.$ Then there exit a pure state $
S_{AB}$ and encodings ${\mathcal E}_{A}^{i}$ such that
\begin{equation}
S_{AB}^{i}=({\mathcal E}_{A}^{i}\otimes {\mathrm Id}_{B})[S_{AB}].  \label{cod}
\end{equation}
\textit{Proof}. For simplicity assume that $S_{B}$ is nondegenerate. Then
\[
S_{B}=\sum_{k=1}^{d}\lambda _{k}|e_{k}^{B}\rangle \langle e_{k}^{B}|,
\]
where $\lambda _{k}>0$ and $\left\{ |e_{k}^{B}\rangle \right\} $ is an
orthonormal basis in ${\mathcal H}_{B}.$ Let $\left\{ |e_{k}^{A}\rangle
\right\} $ be an orthonormal basis in ${\mathcal H}_{A}.$ For a vector $
|\psi^{A}\rangle =\sum_{k=1}^{d}c_{k}|e_{k}^{A}\rangle $ we denote $|\bar{
\psi}^{B}\rangle =\sum_{k=1}^{d}\bar{c}_{k}|e_{k}^{B}\rangle .$ The map $
|\psi^{A}\rangle \rightarrow |\bar{\psi}^{B}\rangle $ is anti-isomorphism of
${\mathcal H}_{A}$ and ${\mathcal H}_{B}.$ Put
\[
|\psi _{AB}\rangle =\sum_{k=1}^{d}\sqrt{\lambda _{k}}|e_{k}^{A}\rangle
\otimes |e_{k}^{B}\rangle ,
\]
so that $S_{AB}=|\psi _{AB}\rangle \langle \psi _{AB}|$ and define encodings
by the relation
\[
{\mathcal E}_{A}^{i}\left[ |\psi^{A}\rangle \langle \phi^{A}|\right] =\langle
\bar{\psi}^{B}|S_{B}^{-1/2}S_{AB}^{i}S_{B}^{-1/2}|\bar{\phi}^{B} \rangle
,\quad |\psi^{A}\rangle ,|\phi^{A}\rangle \in {\mathcal H}_{A}.
\]
Then one can check that ${\mathcal E}_{A}^{i}$ are indeed channels fulfilling
the formula (\ref{cod}).

In the case $S_{B}$ is degenerate, the above construction should be modified
by replacing $S_{B}^{-1/2}S_{AB}^{i}S_{B}^{-1/2}$ in the formula above with $
\sqrt{S_{B}^{-}}S_{AB}^{i}\sqrt{S_{B}^{-}}+P_B^{0}$ where $S_{B}^{-}$ is the
generalized inverse of $S_{B}$ and $P_B^{0}$ is the projection onto the null
subspace of $S_B$.

We now observe an inequality relating the asymptotic
entanglement-assisted and unassisted capacities. Apparently,
\begin{equation}
C_{ea}^{(1)}(\Phi )\le \max_{\pi _{i},{}S_{AB}^{i}}\left[ H\left(
\sum_{i}\pi _{i}(\Phi \otimes {\mathrm Id}_{B})\left[ S_{AB}^{i}\right]
\right) -\sum_{i}\pi _{i}H\left( (\Phi \otimes {\mathrm Id}_{B})\left[
S_{AB}^{i}\right] \right) \right] ,
\end{equation}
where $S_{AB}^{i}$ are already arbitrary states, not necessarily of the form
(\ref{cod}). The quantity on the right hand side is nothing but the one-shot
classical capacity $C^{(1)}(\Phi \otimes {\mathrm Id}_{B})$ of the channel $
\Phi \otimes {\mathrm Id}_{B}.$ It was shown in \cite{shu} that $C^{(1)}(\Phi
\otimes {\mathrm Id}_{B})=C^{(1)}(\Phi )+C^{(1)}({\mathrm Id}
_{B})=C^{(1)}(\Phi )+\log d.$ Applying the same argument to $\Phi ^{\otimes
n}$ instead of $\Phi ,$ we have
\[
C_{ea}^{(1)}(\Phi ^{\otimes n})\le C^{(1)}(\Phi ^{\otimes n})+n\log d.
\]
Dividing by $n$ and taking limit $n\rightarrow \infty ,$ we obtain
\[
C_{ea}(\Phi )\le C(\Phi )+\log d.
\]

One can expect that a similar inequality
\[
C_{ea}(\Phi )\leq C^{(1)}(\Phi )+\log d
\]
holds generally for the one-shot classical capacity; if it breaks for some
channel $\Phi ,$ then for this channel $C^{(1)}(\Phi )<C(\Phi ),$ which
would imply negative answer to the long-standing question concerning
additivity of the classical capacity.

It is not difficult to check that the
inequality indeed holds for all unital qubit channels and for
$d$-depolarizing channel
\begin{equation}\label{dep} \Phi[{\cal S}]=(1-p) S + p \frac{I}{d}
\mbox{Tr} S.  \end{equation}
Here $\dim {{\cal H}}=d$ and the parameter $p$ should lie in the range
$0\leq p \leq \frac{d^2}{d^2-1}$, as  can be seen from the Kraus representation
\begin{equation}\label{rep}
\Phi[{\cal S}]=\left(1-p\frac{d^2-1}{d^2}\right) S+p\frac{1}{d^2}\sum_{\alpha
,\beta\not=d} W_{\alpha \beta } S W_{\alpha \beta }^*, \end{equation}
with $W_{\alpha \beta }; \alpha, \beta=1,\dots,d$ built
upon arbitrary orthonormal basis in ${\cal H}$.

The quantity $C_{ea}(\Phi )$ can be computed by
using unitary covariance of the depolarizing channel and concavity of the
function $S\rightarrow I(S;\Phi)$.  It follows that it achieves the maximum
at the chaotic state $\bar{S}=\frac{I}{d}$. We have
$H(\bar{S})=H(\Phi[\bar{S}])=\log d.$ The entropy exchange $H(\bar{S};\Phi)$
can be computed by as the entropy of the matrix
$[\mbox{Tr}\bar{S}A_{\alpha\beta}^*A_{\alpha\beta}]$, where
$A_{\alpha\beta}=\frac{\sqrt{p}}{d}W_{\alpha\beta}; \alpha,\beta\not=d;
A_{dd}=\sqrt{1-p\frac{d^2-1}{d^2}}I$ are the Kraus operators from the representation (\ref{rep}).
We thus obtain \begin{equation} C_{ea}(\Phi )=\log d^2 + \Big(1-p
\frac{d^2-1}{d^2}\Big)\log \Big(1-p \frac{d^2-1}{d^2}\Big) +p
\frac{d^2-1}{d^2}\log \frac{p}{d^2}.\end{equation} This should be compared
with the unassisted classical capacity, which is equal to
\begin{equation}\label{c1dep}C^{(1)}(\Phi )=\log d +\Big(1-p
\frac{d-1}{d}\Big)\log\Big(1-p \frac{d-1}{d}\Big)+ p
\frac{d-1}{d}\log\frac{p}{d},\end{equation} and is achieved for an
ensemble of equiprobable pure states taken from an orthonormal basis in
${\cal H}$.
One then sees \cite{eac} that
$\frac{C_{ea}(\Phi )}{C^{(1)}(\Phi )}\rightarrow d+1$ in the limit of strong noise
$p\rightarrow 1$ (note that
both capacities tend to zero!)

Moreover, taking the maximal possible value  $p=\frac{d^2}{d^2-1}$, we obtain
$$C_{ea}=\log\frac{d^2}{d^2-1}, $$ $$
C^{(1)}= \frac{1}{d+1}\log\frac{d}{d+1}+\frac{d}{d+1}\log\frac{d^2}{d^2-1}.$$
Here the ratio  $\frac{C_{ea}}{C^{(1)}}$ monotonically increases from the
value 5.0798 for $d=2$, approaching tightly the asymptotic line $2(d+1)$ as
$d$ grows to infinity.

\acknowledgments{The author is grateful to the
authors of the paper \cite{eac2} for making it available for discussion prior
to publication.

This work is partially supported by the grant INTAS 00-738.}

\end{document}